\documentclass[a4paper,12pt]{article}

\usepackage[utf8x]{inputenc}
\usepackage[T1]{fontenc}
\usepackage{times}
\usepackage[colorlinks=true,linkcolor=black,citecolor=black,urlcolor=blue]{hyperref}
\usepackage{geometry}
\usepackage{ragged2e}
\usepackage{authblk}

\usepackage{placeins}

\usepackage{xcolor}

\usepackage{titlesec}

\titleformat{\section}{\bfseries\uppercase}{\thesection.}{1em}{\noindent}

\titleformat{\subsection}{\bfseries}{\thesection.\thesubsection.}{1em}{\noindent}

\usepackage{graphicx} 
\usepackage{multirow} 
\usepackage{cite}
\usepackage{breakurl}
\usepackage{amsmath, amssymb, amsfonts, bm}
\usepackage{txfonts}
\usepackage{enumitem}
\usepackage{xcolor}
\usepackage{enumitem}
\usepackage{booktabs}
\usepackage{tabularx}

\usepackage[flushmargin, bottom]{footmisc}
\usepackage{bigfoot}
\DeclareNewFootnote[para]{default}
\DeclareNewFootnote[para]{B}[fnsymbol]
\MakePerPage{footnote}

\titlespacing*{\subsection}{0pt}{1.5em}{0.2em}
\renewcommand\eqref[1]{Equation~\ref{#1}}
\renewcommand{\thesection}{\arabic{section}}
\renewcommand{\thesubsection}{\arabic{subsection}}

\newlength{\bibitemsep}
\newlength{\bibparskip}
\let\oldthebibliography\thebibliography
\renewcommand\thebibliography[1]{%
  \oldthebibliography{#1}%
  \setlength{\parskip}{\bibitemsep}%
  \setlength{\itemsep}{\bibparskip}%
}

\newcommand{\createtitle}[1]{%
    \begin{justify}%
    \fontsize{16}{20}\selectfont\bfseries#1%
    \end{justify}\vskip0.5cm%
}

\renewenvironment{abstract}{%
    \noindent\textbf{\centerline{ABSTRACT}}\\
    \itshape
}{}

\usepackage{siunitx}
\usepackage{comment}
\usepackage{url}
\usepackage{cleveref}

\Crefname{figure}{Fig.}{Fig.}
\Crefname{table}{Table}{Tables}
\Crefname{section}{Section}{Sections}

\newcommand{\citespl}{{\cite{Watcharasupat2022AutonomousGain}}}

\begin{document}

\begin{center}
	\includegraphics[width=2.65in]{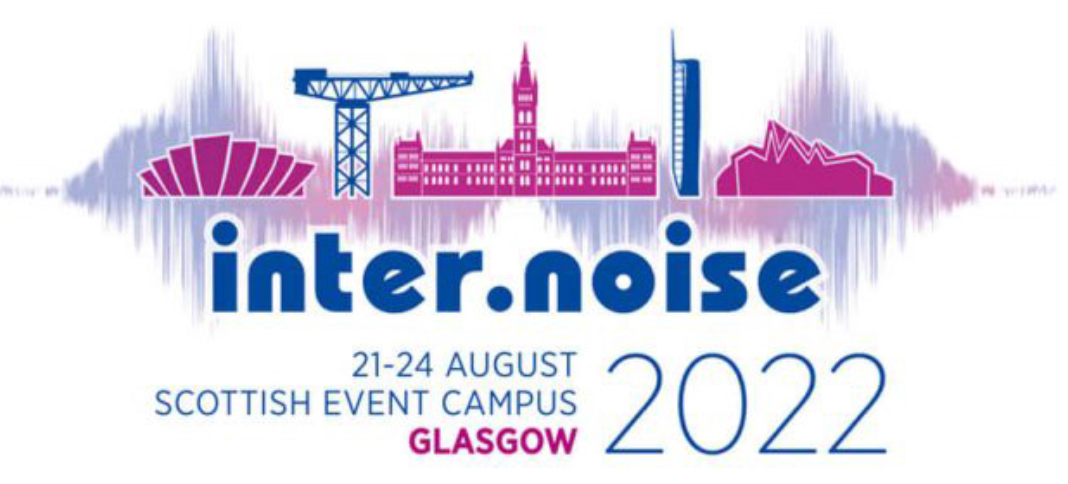}
\end{center}
\vskip.5cm


\createtitle{Deployment of an IoT System for Adaptive In-Situ Soundscape Augmentation}
\renewcommand\baselinestretch{1}
\begin{flushleft}
Trevor~Wong\footnote{trevor.wong@ntu.edu.sg}\textsuperscript{,}\footnoteB[1]{Equal contribution}, 
Karn~N.~Watcharasupat\footnote{karn001@e.ntu.edu.sg}\textsuperscript{,}\footnotemarkB[1],
Bhan~Lam\footnote{bhanlam@ntu.edu.sg}, 
Kenneth~Ooi\footnote{wooi002@e.ntu.edu.sg}, 
Zhen-Ting~Ong\footnote{ztong@ntu.edu.sg},
Furi~Andi~Karnapi\footnote{furi\_andi\_karnapi@tech.gov.sg}\textsuperscript{,}\footnoteB[3]{F. A. Karnapi is currently with the Government Technology Agency, Singapore. His contribution to this work was done while he was with Nanyang Technological University,  Singapore.},  Woon-Seng~Gan\footnote{ewsgan@ntu.edu.sg}

\vspace{.5\baselineskip}
Digital Signal Processing Laboratory, School of Electrical and Electronic Engineering,\\
Nanyang Technological University,
50 Nanyang Avenue, S2-B4a-03, Singapore 639798

\vspace{\baselineskip}


\end{flushleft}





\begin{abstract}
    Soundscape augmentation is an emerging approach for noise mitigation by introducing additional sounds known as ``maskers'' to increase acoustic comfort. Traditionally, the choice of maskers is often predicated on expert guidance or post-hoc analysis which can be time-consuming and sometimes arbitrary. Moreover, this often results in a static set of maskers that are inflexible to the dynamic nature of real-world acoustic environments. Overcoming the inflexibility of traditional soundscape augmentation is two-fold. First, given a snapshot of a soundscape, the system must be able to select an optimal masker without human supervision. Second, the system must also be able to react to changes in the acoustic environment with near real-time latency. In this work, we harness the combined prowess of cloud computing and the Internet of Things (IoT) to allow in-situ listening and playback using microcontrollers while delegating computationally expensive inference tasks to the cloud. In particular, a serverless cloud architecture was used for inference, ensuring near real-time latency and scalability without the need to provision computing resources. A working prototype of the system is currently being deployed in a public area experiencing high traffic noise, as well as undergoing public evaluation for future improvements.
    %
\end{abstract}



\section{Introduction}


\textit{Soundscape}, as defined in the ISO 12913-1 standard, is an ``acoustic environment as perceived or experienced and/or understood by a person or people, in context'' \cite{InternationalOrganizationforStandardization2014ISOFramework}. The ISO 12913 series of standards \cite{InternationalOrganizationforStandardization2014ISOFramework, InternationalOrganizationforStandardization2018ISO/TSRequirements, InternationalOrganizationforStandardization2019ISO/TSAnalysis} represents a paradigm shift in sound environment management and details perceptual methods to holistically assess and analyse the sound environment. As opposed to traditional noise management, the soundscape management framework perceives sound as a resource rather than a waste \cite{Pijanowski2011}; focuses on sounds of preference rather than sounds of discomfort; and manages unwanted with wanted sounds as well as reducing unwanted sounds rather than just reducing sound levels \cite{Kang2016TenEnvironment}. Hence, there are soundscape intervention techniques based on augmentation or introduction of \textit{masking} sounds into the acoustic environment to improve the overall perception of acoustic comfort. Commonly, such interventions involve augmentation with natural sounds (e.g., via loudspeakers) and have been trialled in outdoor recreational spaces \cite{VanRenterghem2020InteractivePark, Hong2021ASounds}, and indoors, such as in offices \cite{Abdalrahman2020Audio-visualOffices} and nursing homes \cite{Devos2018SoundscapeDementia}, to improve acoustic comfort. 

\subsection{Soundscape augmentation}

Augmentation of soundscapes with wanted sounds is akin to sound masking systems (such as the one described by Eriksson et al. in \cite{Eriksson2017TheSpace}), which are commonly employed in office environments to reduce distractions. The key difference between sound masking systems and the soundscape-based augmentation systems is that masking is based on objective metrics, whereas soundscape augmentation is based largely on subjective metrics rooted in context. The context ``includes the interrelationships between person and activity and place, in space and time'' and is further detailed with examples in ISO 12913-1 \cite{InternationalOrganizationforStandardization2014ISOFramework}. Soundscapes are also not limited to real-world environments, and can encompass virtual environments and even environments recalled from memory. 

Until recently, state-of-the-art methods for in-situ soundscape augmentation or masking interventions involved static playback of sounds through a loudspeaker, wherein the sound levels and tracks have to be manually selected either with the assistance of experts or via listening tests \cite{VanRenterghem2020InteractivePark, Hong2021ASounds}, limiting the diversity of possible masker configurations resulting system inflexibility in reacting to changes in the acoustic environment. A recent work by Ooi et al.\ \cite{Ooi2022ProbablyAugmentation} was among the first to utilize a deep learning model termed a ``probabilistic perceptual attribute predictor'' (PPAP) to automatically select a masker track that yields the most optimal value on a predefined subjective metric given a short recording of the ambient soundscape, allowing near-realtime autonomous masker selection. The model in \cite{Ooi2022ProbablyAugmentation} was trained on a large dataset of human subjective responses collected in line with the ISO 12913-2 standard \cite{InternationalOrganizationforStandardization2018ISO/TSRequirements}. The model in \cite{Ooi2022ProbablyAugmentation}, however, only outputs the optimal masker track but not the gain level, thus requiring the same masker track with different playback gain levels to be treated as separate tracks. A follow-up work in \citespl{} incorporated the masker gain level explicitly into the model, in addition to reformulating the input-output structure, allowing for a more optimized querying process for in-situ deployment.

\subsection{In-situ Augmentation with internet-of-things (IoT)}

Given a predictive model for soundscape augmentation, we therefore aim to deploy it as part of a real-world soundscape augmentation system, using an on-site Internet-of-Things (IoT) system for ambient sensing and playback with a cloud inference engine. In this work, Amazon Web Services (AWS) \cite{AmazonWebServices2022OverviewWhitepaper} was used as the cloud service provider.

\begin{figure}[t]
    \centering
    \includegraphics[width=0.8\textwidth]{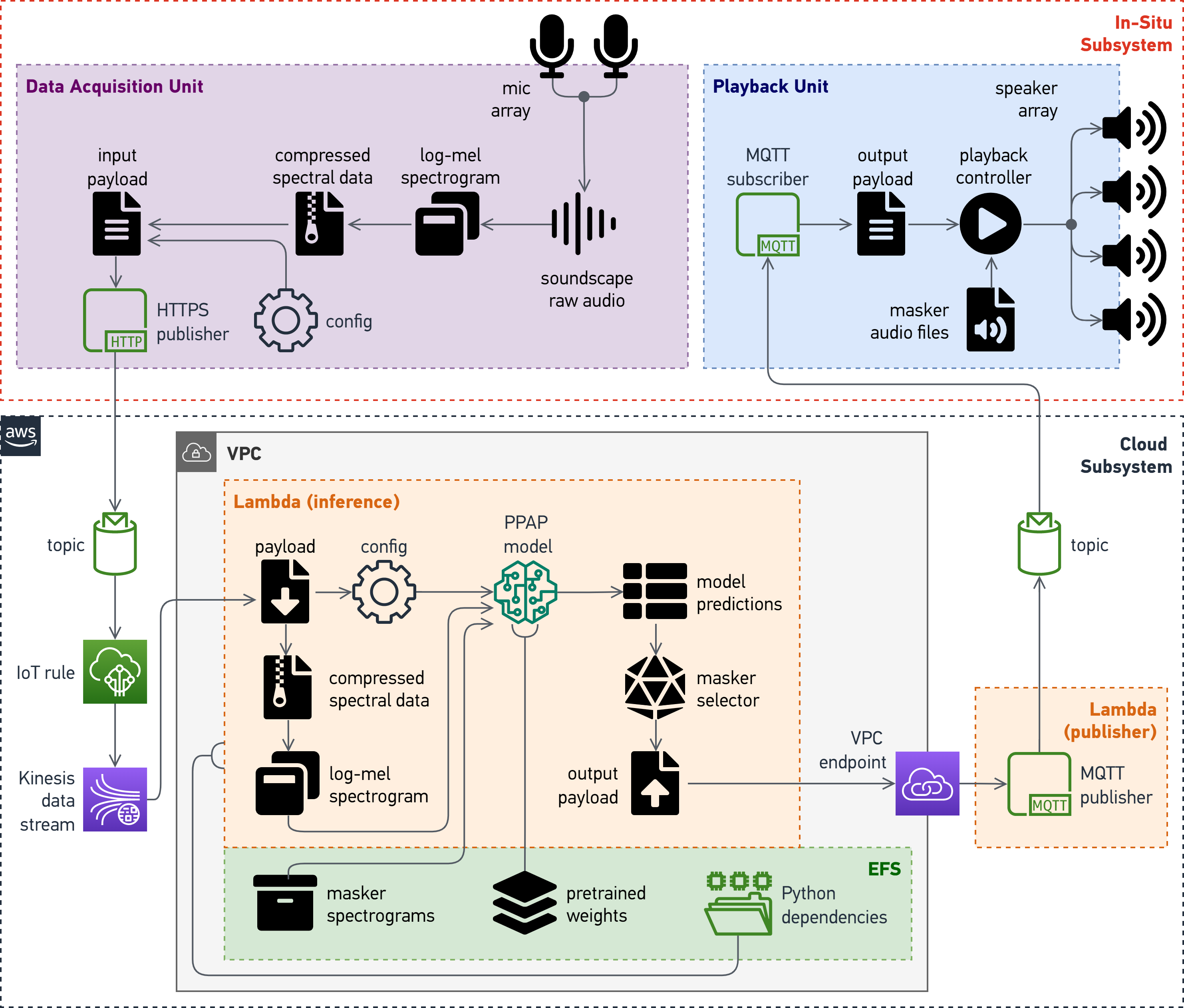}
    \caption{Overview of the soundscape augmentation system.}
    \small VPC: Virtual Private Cloud. EFS: Elastic File System.
    \label{fig:overview}
\end{figure}

The proposed soundscape augmentation system consists of two main subsystems as shown in \Cref{fig:overview}, namely, the in-situ subsystem and the cloud subsystem. The in-situ subsystem itself consists of two units: the data acquisition unit and the playback unit. At the top level, the data acquisition unit is responsible for recording and transmitting the ambient soundscape audio data to the cloud subsystem. The cloud subsystem then passes the received audio data to the inference engine and sends the prediction outputs back to the playback unit within the in-situ subsystem. Back at the installation site, the playback unit receives the payload, and outputs the optimal masker track through the speakers at the specified gain level accordingly. 

The following sections are organized as follows. \Cref{s:acq} discusses the in-situ data acquisition Unit, \Cref{s:cloud} discusses the cloud system, and \Cref{s:playback} discusses the playback unit of the on-site subsystem. Finally, \Cref{s:conclusion} concludes the paper.

\section{Data Acquisition}\label{s:acq}

At the installation site, raw audio is continuously captured at a sampling frequency of \SI{44.1}{\kilo\hertz} via a miniDSP UMA-8 microphone array\footnote{https://www.minidsp.com/products/usb-audio-interface/uma-8-microphone-array} connected to a Raspberry Pi 3 Model B+ microcontroller\footnote{https://www.raspberrypi.com/products/raspberry-pi-3-model-b-plus/}, similar to the deployment unit used in \cite{Tan2021ExtractingSystem}. Although the UMA-8 array has seven onboard MEMS microphones, we only require two of the channels for this work. The two-channel raw audio is then converted into a log-mel spectrogram and compressed using a technique described in Section \ref{s:acq}.\ref{s:acq/compression} to reduce the overall packet size and egress rate requirement. The compressed data together with additional metadata is then encoded and published via HTTPS to an AWS IoT endpoint. 

\begin{table}[tb]
    \caption{Egress rate requirement for different audio data representations (without compression). For this paper, sampling rate $f_\text{s}=\SI{44.1}{\kilo\hertz}$, number of channel $C=2$, STFT frame size $L=4096$, STFT hop size $R=2048$, and mel bins $M=64$.}
    \vspace{\baselineskip}
    \centering
    \begin{tabular}{lS[table-format=2.0]S[table-format=2.2]S[table-format=3.2]S[table-format=5.0]}
    \toprule
    &&\multicolumn{3}{c}{\shortstack{Egress Rate}}\\
    \cmidrule(lr){3-5}
        \multicolumn{1}{l}{Representation} 
        & \multicolumn{1}{r}{Bit Depth} 
        & \multicolumn{1}{r}{kHz}
        & \multicolumn{1}{r}{\si{kB\per\second}}
        & \multicolumn{1}{r}{\si{kB\per(\SI{30}{\second})}}\\
    \midrule
        Raw audio 
            & 16 & 88.20 & 172.27 & 5168 \\
        \midrule
        Linear-frequency spectrogram 
            & 32 & 88.24 & 344.70 & 10341\\
            & 16 & 88.24 & 172.35 & 5170 \\
        \midrule
        Mel-frequency spectrogram         
            & 32 & 2.76 & 10.77  & 323\\
            & 16 & 2.76 & 5.38   & 161\\
    \bottomrule
    \end{tabular}
    \label{tab:egress}
\end{table}

One of the key obstacles to overcome was the high amount of data usage for transmission of the audio data from the edge and the cloud. As the prediction model requires \SI{30}{\second} of audio data per frame, streaming the raw audio to the cloud would require a \SI{172.27}{kB\per\second} egress rate (see \Cref{tab:egress}), which not only incurs significant network cost in the long run but also increases the overall latency of the system. In contrast, since the prediction model only requires a mel spectrogram input, streaming the transformed data instead of the raw audio would only require a \SI{5.38}{ kB\per\second} egress rate at 16-bit precision.

Even with the reduced egress rate, however, \SI{30}{\second} of mel spectrogram at 16-bit precision would still require \SI{161}{kB} of data per payload. This exceeds the \SI{128}{kB} payload limit of the AWS IoT message broker \cite[pp.~347]{AmazonWebServices2022AWSGuide}. Thus, it was necessary to further compress the spectral data to meet the payload requirement.
\FloatBarrier
\subsection{Compression Technique} \label{s:acq/compression}

\begin{figure}
    \centering
    \includegraphics[width=\textwidth]{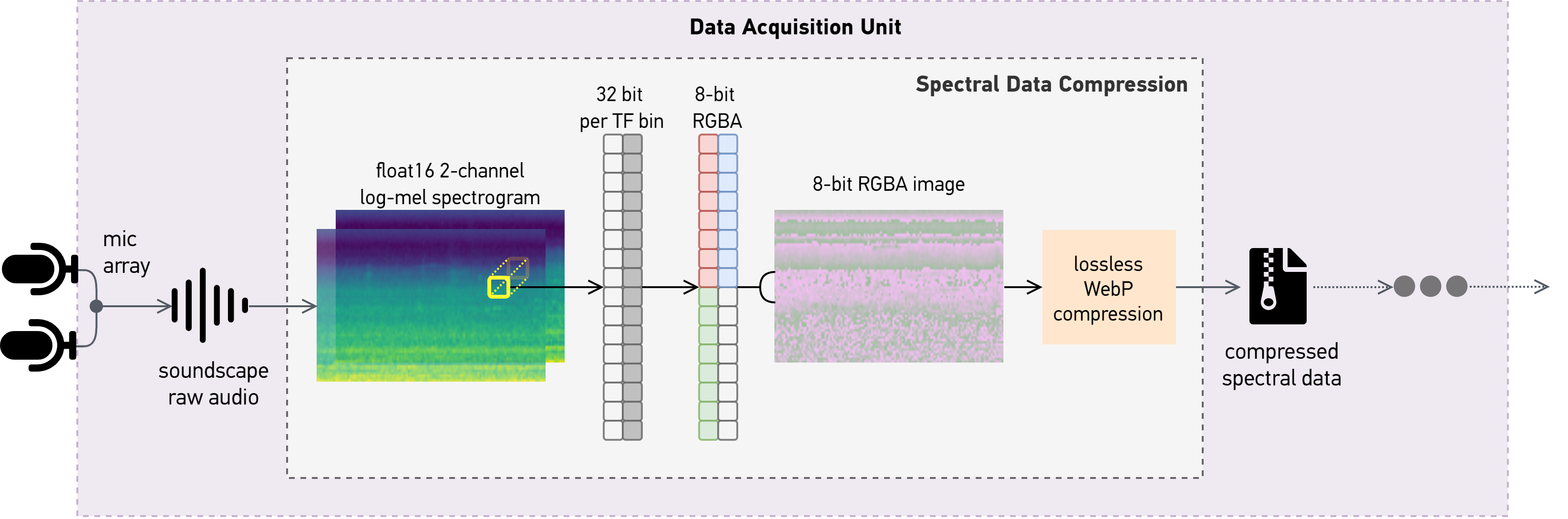}
    \vspace{-\baselineskip}
    \caption{Proposed compression technique for floating-point spectral data.}
    \label{fig:compression}
\end{figure}

The overview of the compression method is shown in \Cref{fig:compression}. Each time-frequency (TF) bin of the mel spectrogram consists of 2 channels of 16-bit floating-point numbers, totalling 32 bits. Conveniently, RGBA representation of image at 8-bit color depth also uses 32 bits per pixel. Exploiting the availability of lossless image compression, we regrouped the 32 bits of spectral data into 4 `channels' of 8 bits. Each of these 8-bit groups can then be viewed as one RGBA pixel. Applying this representation to the entire spectrogram allows the 2-channel mel spectrogram to be viewed as an 8-bit RGBA image. As an `image', a lossless WebP compression\footnote{https://developers.google.com/speed/webp} \cite{Alakuijala2012WebPSpecification} can be applied to create a compressed image with a size typically less than \SI{100}{kB}. 

\subsection{Payload transmission} \label{s:acq/tocloud}

The compressed image is then viewed as a bitstream to form part of the dictionary-based payload that is then encoded into a text-based format using the ISO/IEC 8859-1 (\texttt{latin-1}) specification \cite{InternationalOrganizationforStandardization1998ISO/IEC1} before being published to the endpoint using the AWS IoT Core message broker. The AWS IoT message broker supports both HTTPS and MQTT protocols for publishing. We initially attempted to transmit the spectral data from the Raspberry Pi to the cloud using the lighter MQTT protocol. However, since AWS only supports a Quality of Service (QoS) of 0 or 1, we found that transmission is somewhat unstable. In contrast, HTTPS runs over a reliable transport layer (typically TCP/IP) which allows a more stable transmission. As the Raspberry Pi can handle HTTPS overhead without much additional delay compared to MQTT, we decided to use HTTPS publishing for the audio data.

\section{Cloud subsystem} \label{s:cloud}

Once the data is published to the MQTT topic, an AWS IoT Rule forwards the data to a Kinesis data stream. A Python script housed in AWS Lambda accesses the data in the Kinesis stream and decodes the compressed payload to obtain the log-mel spectrogram. The spectrogram is then passed to the cloud model which outputs predictions for the most suitable maskers and the gain at which they should be played based on the input log-mel spectrogram. Any model that takes log-mel spectrograms as input and outputs predictions of most suitable maskers and gain levels can be used as the cloud model, but for illustration purposes, we have used the PPAP model as shown in Fig. \ref{fig:overview}. The JSON-formatted predictions are then forwarded to another Lambda outside the Virtual Private Cloud (VPC) which then publishes the data to a separate MQTT topic subscribed by the Playback Unit on-site.


\subsection{Serverless engine} \label{s:cloud/serverless}
The inference stage is handled in a serverless style which offers a more cost-effective deployment option that can easily scale with the number of in-situ units. Crucially, only one cloud pipeline is needed for any number of in-situ installations. As the number of in-situ units increases, the load balancing and scaling are taken care of by the Lambda concurrency which automatically creates multiple instances of the same inference engine to cope with the load.

Since the Lambda environment only supports ephemeral storage, repopulating the environment during initialization of the execution environment, with the required deep learning dependencies, pretrained model weights, and a masker bank would significantly increase the cold start time and overall latency. Moreover, the deployment package has a strict limit of \SI{250}{MB} and, prior to late March 2022\footnote{In late March 2022, the ephemeral storage limit of AWS Lambda has been increased from \SI{512}{MB} to \SI{10}{GB}.}, the temporary storage had a strict limit of \SI{512}{MB} \cite[pp.~443--444]{AmazonWebServices2022AWSGuide}, which are insufficient for the TensorFlow \cite{Abadi2016TensorFlow:Learning} and NumPy \cite{Harris2020ArrayNumPy} dependencies. As a result, we utilized the Amazon Elastic File System (EFS) to store the required data and dependencies, and configure the Lambda function to dynamically mount the shared file system during the cold start phase. The file system was prepared using AWS Code Build. Note that it is also possible to use a container image for Lambda deployment, which would allow up to \SI{10}{GB} of container size. For this work, the deployment package option was used due to the lower development complexity required. 

\subsection{Automatic masker selection} \label{s:cloud/amss}

The masker selection is performed by a pretrained model from \citespl{}. The model takes three inputs, namely, the log-mel spectrogram of the ambient soundscape, the candidate maskers, and the candidate gain levels for each masker. In this work, the spectrograms of the 200 masker candidates are already precomputed and stored in the mounted EFS.

For each masker-gain pair, the PPAP model outputs a probability distribution of the predicted pleasantness, parameterized by the mean and the log-standard-deviation. Since exhaustive inference of all possible gain values would result in unnecessarily many candidate masker-gain pairs, each masker is paired with 5 random gain values drawn from a log-normal distribution with a mean of $-2.0$ and standard deviation of $1.5$. The parameters of the gain distribution were obtained from an exhaustive calibration of the masker banks with five soundscape-to-masker ratios of \num{-6}, \num{-3}, \num{0}, \num{3}, \num{6} dBA against the soundscape recordings in the Urban Soundscape of the World dataset \cite{DeCoensel2017UrbanMind}, using the automated calibration method in \cite{Ooi2021AutomationHead}. In practice, the gain distribution can be further refined to better suit the acoustic characteristics of the deployment site.

The pretrained model and log-mel spectrograms of the candidate maskers, which are precomputed and stored in the file system, are already loaded into the Lambda memory during the cold start. Upon invocation, the decoded spectrogram of the soundscape is passed to the model along with the maskers and the gains to obtain the predictions.

\FloatBarrier
\subsection{Cost-Speed Optimization} \label{s:cloud/tuning}

\begin{figure}[hbt]
    \centering
    \includegraphics[width=0.8\textwidth]{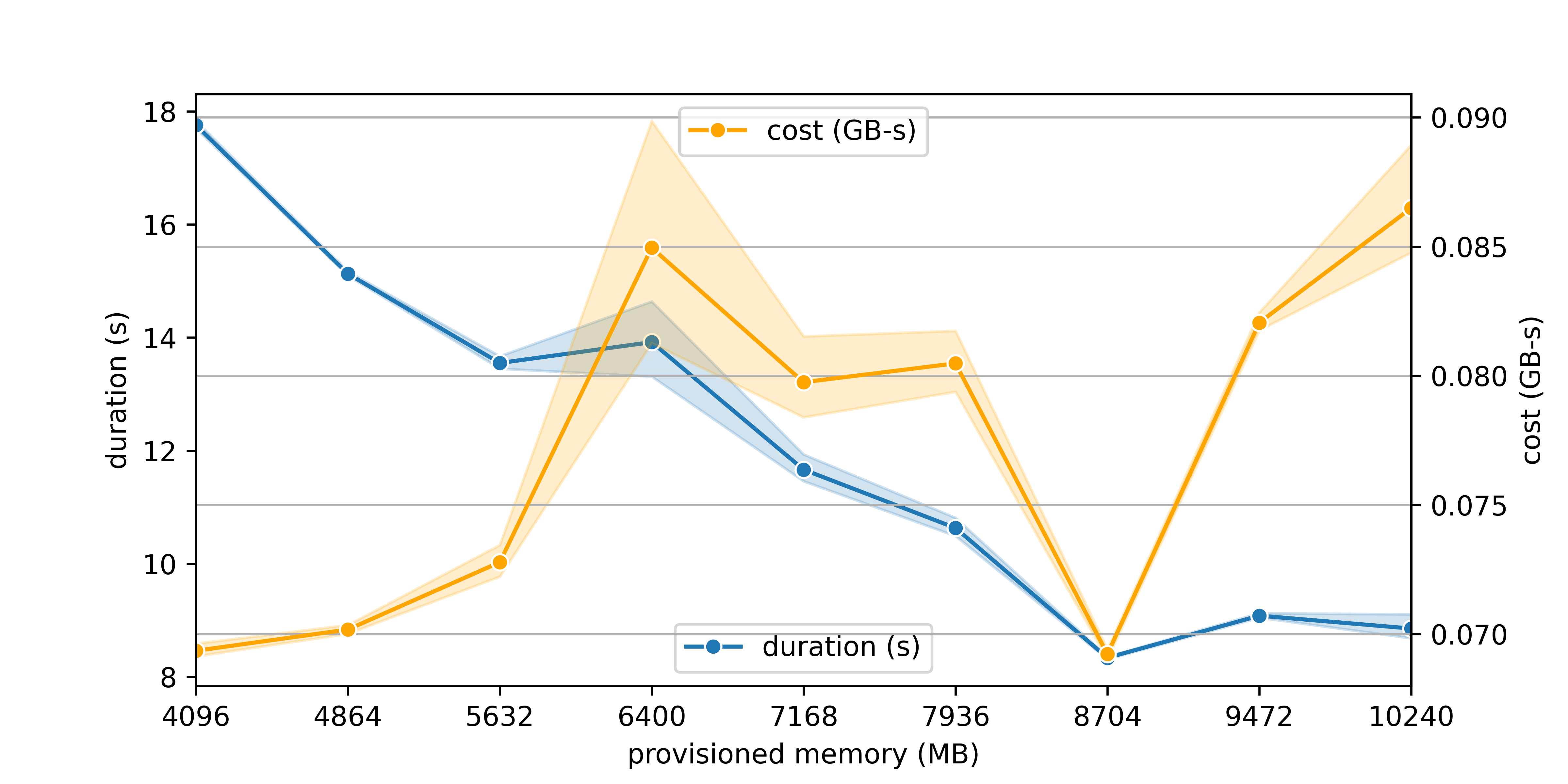}
    \caption{Results of the AWS Lambda Power Tuning cost-speed optimization}
    \label{fig:AWSpowertuning}
\end{figure}

The serverless nature of Lambda allows the provisioned memory to be configured between \SI{128}{MB} and \SI{10}{GB}. Higher provisioned memory will generally improve the speed of the Lambda functions but will incur higher costs per unit runtime. In order to optimize the cost-effectiveness, we utilized AWS Lambda Power Tuning \cite{Casalboni2022AWSTuning} to adjust the configured memory. We tested the inference Lambda function at eight memory values between \SI{4}{GB} to \SI{10}{GB} over 20 runs, excluding the cold start run. The results of the tuning runs are shown in \Cref{fig:AWSpowertuning}. Based on the results, \SI{8704}{MB} of provisioned memory provided the fastest mean runtime at \SI{8.34}{\second} and \SI{0.0692}{GB-s}.

\subsection{Prediction Streaming}

Once the PPAP model has predicted the pleasantness of the masker-gain pairs for the input soundscape, the best five pairs are selected. The identifier of each masker and the corresponding gain levels are then packaged into a dictionary that is forwarded via a VPC endpoint to another Lambda function outside the VPC. At the publisher Lambda, the payload is received and forwarded to another MQTT topic subscribed to by the playback unit in the in-situ subsystem.

The forwarding via a VPC endpoint is done for two reasons. First, VPCs do not have internet access by default. Connecting a VPC to the internet requires a Network Address Translation (NAT) gateway, which is a relatively expensive service. Directly publishing from the inference Lambda to the MQTT topic would thus require a NAT, resulting in unnecessarily high service cost. Second, MQTT publishing requires very low computational resources. Moving the publishing operation out of the inference Lambda, which has high provisioned memory, reduces the overall cost, since the publisher Lambda outside the VPC can be provisioned with minimal resources. 

\subsection{Monitoring Dashboard}
\begin{figure}[tb]
    \centering
    \includegraphics[width=\textwidth]{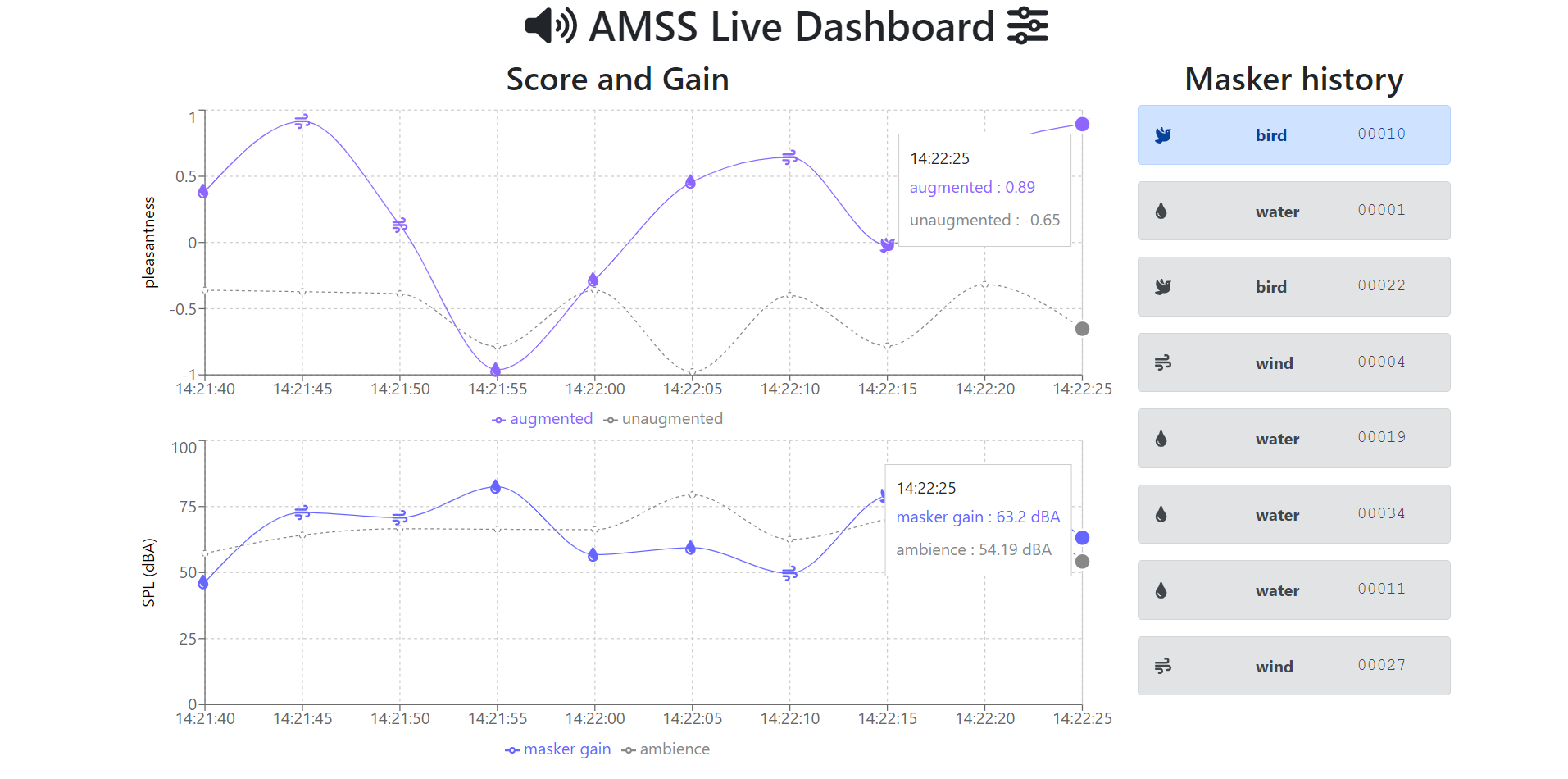}
    \vspace{-1cm}
    \caption{Monitoring dashboard for the soundscape augmentation system}
    \label{fig:monitor}
\end{figure}

A web-based live monitoring dashboard was created using AWS Amplify for convenient monitoring of the system, without the need to access the cloud logs. The dashboard subscribes to the same MQTT topic as the playback unit, allowing it to access the masker-gain pairs recommended by the PPAP model. A sample screenshot of the dashboard is shown in \Cref{fig:monitor}.

\section{Playback and Augmentation} \label{s:playback}

At the installation site, a second Raspberry Pi subscribes to the MQTT topic to receive the prediction payloads from the cloud engine. This Raspberry Pi then continuously outputs the presently most suitable masker at the corresponding gain level through a setup of four full-range speakers to complete the augmentation process.

\subsection{Playback}
The playback Raspberry Pi\footnote{In our work, the playback controller is a Raspberry Pi 4 Model B. Raspberry Pi 3 Model B or B+ can be used as well.}\textsuperscript{,}\footnote{ https://www.raspberrypi.com/products/raspberry-pi-4-model-b} utilizes a single audio stream created by the Python \texttt{sounddevice} library\footnote{https://github.com/spatialaudio/python-sounddevice} to playback masker tracks. Once the predictions are received by the playback Raspberry Pi, the system checks if the top-ranked prediction has either a different masker or substantially different gain from the masker that is currently playing. If either of these conditions is met, the system will begin playing the new masker at the recommended gain level. To prevent the change of masker from being too perceptually jarring, crossfading is implemented when changing to a new masker. If no change in masker or gain is recommended by the model before the current masker track ends, the system will play the same masker again from the start in a simple loop.

The playback Raspberry Pi is connected to a miniDSP MCHstreamer\footnote{https://www.minidsp.com/products/usb-audio-interface/mchstreamer} which outputs the audio signal in a four-channel I2S format. The I2S audio is then converted to a four-channel line-level analog signal using two Adafruit UDA1334A stereo decoders\footnote{https://www.adafruit.com/product/3678}. The analog signal is then amplified by a TDA7850 four-channel amplifier\footnote{https://www.st.com/en/automotive-infotainment-and-telematics/tda7850.html} and sent to four full-range speakers with subwoofers and low-pass crossover filters. 


\section{Conclusion} \label{s:conclusion}
In this work, we have presented an IoT-based soundscape augmentation system that leverages the AWS cloud architecture to perform machine inference tasks in tandem with in-situ microcontroller-based acquisition and playback units. Additionally, we developed a spectral data compression technique exploiting an existing lossless image compression technique. The proposed system is one of the first utilizing a deep learning model for in-situ soundscape augmentation that can autonomously react to changes in the acoustic environment. Public feedback systems \cite{Karnapi2021DevelopmentEvaluation} can be installed at the real-life installation site to access the efficacy of the proposed system, and to make future improvements as part of a co-creation process with users of the soundscape augmentation system.

\section*{Acknowledgements}
\noindent
This work was supported by the National Research Foundation, Singapore, and Ministry of National Development, Singapore under the Cities of Tomorrow R\&D Program (CoT Award: \mbox{COT-V4-2020-2}). Any opinions, findings and conclusions or recommendations expressed in this material are those of the authors and do not reflect the view of National Research Foundation, Singapore, and Ministry of National Development, Singapore. This work was additionally supported by the AWS Singapore Cloud Innovation Center.


\bibliographystyle{ieeetr}
\bibliography{references,extra} 

\end{document}